\documentclass[carbon,aricle,accept,pdftex,moreauthors]{Definitions/mdpi}

\graphicspath{{Figures/}}
\usepackage{amsmath}
\usepackage{amsfonts}
\usepackage{amssymb,bm}
\usepackage{siunitx}
\usepackage{color}
\usepackage{braket}
\usepackage{graphics}

\firstpage{1}
\pubvolume{9}
\issuenum{3}
\articlenumber{84}
\pubyear{2023}
\copyrightyear{2023}
\externaleditor{Academic Editors: Ahmet Sinan Oktem and Matteo Strozzi}
\datereceived{5 July 2023}
\daterevised{24 August 2023} 
\dateaccepted{27 August 2023}
\datepublished{31 August 2023 }
\hreflink{https://doi.org/10.3390/c9030084} 

\Title{Large-Separation Behavior of the Casimir--Polder Force from Real Graphene
Sheet Deposited on a Dielectric Substrate}

\TitleCitation{Large-\\Separation Behavior of the Casimir--Polder Force from Real Graphene
Sheet Deposited on a Dielectric Substrate}

\Author{{{Galina} 
 L. Klimchitskaya} $^{1,2,}$*\orcidA{}
and
Vladimir M. Mostepanenko ${}^{1,2,3}$\orcidB{}}

\AuthorNames{Galina L. Klimchitskaya and Vladimir M. Mostepanenko}

\AuthorCitation{Klimchitskaya, G.L.; Mostepanenko, V.M.}

\address{%
$^{1}$ \quad Central Astronomical Observatory at Pulkovo of the Russian Academy of Sciences, \mbox{196140 Saint Petersburg, Russia;} {vmostepa@gmail.com} 
\\
$^{2}$ \quad {Peter} 
 the Great Saint Petersburg Polytechnic University, 195251 Saint Petersburg, Russia\\
$^{3}$ \quad {Kazan} 
 Federal University, 420008 Kazan, Russia}

\corres{Correspondence: g.klimchitskaya@gmail.com}

\abstract{The Casimir--Polder force between atoms or nanoparticles and graphene-coated dielectric substrates
is investigated in the region of large separations. Graphene coating with any value of the energy gap
and chemical potential is described in the framework of the Dirac model using the formalism of the
polarization tensor. It is shown that the Casimir--Polder force from a graphene-coated substrate
reaches the limit of large separations at approximately 5.6~$\upmu$m distance between an atom or
a nanoparticle and graphene coating independently {{of}} the values of the energy gap and chemical
potential. According to our results, however, the classical limit, where the Casimir--Polder force no
longer depends on the Planck constant and the speed of light, may be attained at much larger
separations depending on the values of the energy gap and chemical potential. In addition, we have
found a simple analytic expression for the Casimir--Polder force from a graphene-coated substrate
at large separations and determined the region of its applicability. It is demonstrated that the
asymptotic results for the large-separation Casimir--Polder force from a graphene-coated
substrate are in better agreement with the results of numerical computations for the graphene
sheets with larger chemical potential and smaller energy gap. Possible applications of the
obtained results in nanotechnology and bioelectronics are discussed. }


\begin{document}

\section{Introduction}

The Casimir--Polder force is a typical example of dispersion interactions caused by a fluctuating
electromagnetic field which is zero in the mean but possesses a nonzero dispersion. Discovered
by Casimir and Polder~\cite{1} as a relativistic generalization of the van der Waals interaction
between two atoms (molecules) or an atom and a material surface, it has received wide
recognition in both fundamental physics dealing with precision experiments and in
nanotechnology (for more, see the monographs~\cite{2,3,4,5,6,7}).

The modern theoretical description of the Casimir--Polder force is based on quantum electrodynamics.
In the original article~\cite{1}, Casimir and Polder considered the interaction of two neutral
atoms and an atom with an ideal metal plate. The~case of an atom spaced at large separation
from a plate made of real material characterized by the frequency-dependent dielectric
permittivity was considered in~\cite{8}. This was done on the basis of the Lifshitz theory~\cite{9}
describing the Casimir force acting between two thick plates by rarefying the material of one
of~them.

A more general formula for the potential energy of a molecule of solute in a dilute solution
of low concentration as a function of the interface distance was obtained in~\cite{10} using
the same rarefying procedure of one of the plates (see also~\cite{10a}). At~a later time, the~Lifshitz
formula for the Casimir--Polder interaction of microparticles with material plates was
used in several hundreds papers (see~\cite{2,3,4,5,6,7} and references therein). This formula
allows calculation of the Casimir--Polder interaction given the dynamic atomic (molecular)
polarizability and the dielectric permittivity of plate material. Formulas of this kind are
applicable in the case of nanoparticles interacting with planar or gently curved surfaces
under the condition that the nanoparticle radius $R$ is much less than the distance to the
surface~\cite{11,12,13,13a}.

Among the new materials which have appeared, the~one-atom-thick layer of carbon atoms
called graphene seems to hold the lead~\cite{14}. Thanks to its two-dimensional structure,
graphene admits a theoretical description in the framework of the Dirac formalism of quantum
electrodynamics~\cite{15,16} without resorting to the phenomenological assumptions and
far-reaching extrapolations which are often used in condensed matter physics when dealing with
ordinary three-dimensional materials. This enables the response of graphene to the
electromagnetic field to be found as a function of frequency, wave vector, and temperature
with the use of the polarization tensor~\cite{17,18,19,20}. In~actual fact, the~two independent
components of the polarization tensor are equivalent to the spatially nonlocal dielectric
permittivities of graphene~\cite{21,22}. Because~of this, by using this formalism one can calculate
the Casimir--Polder force acting from the source side of graphene on atoms, molecules, and
nanoparticles.

\textls[-15]{Calculations of the Casimir--Polder force from graphene acting on atoms (see, e.g.,~\mbox{\cite{23,24,25,26,27,28,29,30,31,32}})} and nanoparticles (see, e.g.,
\cite{32a,33,34,35,36,37,38,39}) find applications in both fundamental physics and
nanotechnology. An~expression for the force takes an especially simple form in the limiting case
of large separations (high temperatures). At~room temperature $T$ = 300 K, this is reached at
a few micrometers. In~\cite{40}, the~large separation behavior of the Casimir--Polder force
was investigated for a graphene sheet with no foreign atoms possessing zero chemical
potential and sufficiently small energy gap in the spectrum of~quasiparticles.

Real graphene sheets are characterized by nonzero values of both the energy gap
$\Delta$ and chemical potential $\mu$.  In~a recent article~\cite{41}, the~large-separation
behavior of the Casimir--Polder force was explored for a real graphene
sheet freestanding in vacuum as a function of $\Delta$ and $\mu$.
It was shown that it is possible to control the force
value by varying the values of $\Delta$ and $\mu$ and deal with the analytic asymptotic
expressions for the force rather than with more complicated computations by means of
the Lifshitz~formula.

In practical implementations, graphene is usually deposited on a dielectric substrate. Here,
we investigate the large-separation behavior of the Casimir--Polder force acting on atoms or
nanoparticles from real graphene sheets characterized by an energy gap and
chemical potential which are deposited on thick silica glass substrate. We find that
the range of large separations for a graphene-coated substrate starts at approximately
5.6 $\upmu$m from the surface and is almost independent of the values of
$\Delta$ and $\mu$. The~ratio of forces from a graphene-coated substrate and an
uncoated silica glass plate decreases with increasing $\Delta$ and increases with
increasing $\mu$. It is shown that the classical limit of the Casimir--Polder force from
an ideal metal plane reached at the thermal length, for~a graphene-coated substrate
may be reached at much larger separations from the surface which essentially depend
on the values of $\Delta$ and $\mu$.

The analytic asymptotic expression for the large-separation behavior of the Casimir--Polder
force from a graphene-coated substrate is found as a function of separation from the
surface, energy gap, chemical potential, and~temperature. The~asymptotic results are
compared with numerical computations of the Casimir--Polder force at large separations
for different values of $\Delta$ and $\mu$. According to our results, at~some fixed
separation an~agreement between the asymptotic and numerical results becomes better
with increasing chemical potential and worse with increasing energy gap. By~and {{large}},
for a graphene-coated substrate the asymptotic results are found to be in slightly better
agreement with the results of numerical computations than for a freestanding graphene~sheet.

The structure of the article is as follows. Section~\ref{sec:2} is devoted to the general formalism of
the Lifshitz theory for graphene deposited on a substrate. In~Section~\ref{sec:3}, we consider the
impact of a substrate on the large-separation Casimir--Polder force from real graphene sheet.
Section~\ref{sec:4} is devoted to a confrontation between the analytic asymptotic results and
numerical computations in the presence of a substrate. Section~\ref{sec:5} provides
a discussion, while Section~\ref{sec:6} lists our~conclusions.

\section{General~Formalism}\label{sec:2}
\newcommand{\Mr}{{r_{\rm TM}(i\zeta_l,y)}}
\newcommand{\Er}{{r_{\rm TE}(i\zeta_l,y)}}
\newcommand{\Fv}{{\mbox{$\tilde{v}_F$}}}
\newcommand{\Sv}{{\mbox{$\tilde{v}_F^2$}}}
\newcommand{\tP}{{\mbox{$\widetilde{\Pi}$}}}
\newcommand{\kbot}{{k_{\bot}}}
\newcommand{\ve}{{\varepsilon}}

Here, we briefly present the formalism of the Lifshitz theory describing the
Casimir--Polder force between an atom or a nanoparticle and any planar structure,
e.g., a~graphene-coated substrate, which are at temperature $T$ and separated
by a distance $a$. Using the dimensionless Matsubara frequencies
$\zeta_l=2a\xi_l/c=4\pi ak_BTl/(\hbar c)$, where $k_B$ is the Boltzmann constant,
$l=0,1,2,\dots$, and $y=(4a^2k_{\bot}^2+\zeta_l^2)^{1/2}$ {with} 
 $k_{\bot}=|\bf k_{\bot}|$,
$\bf k_{\bot}$ being the projection of the wave vector on the planar structure, the~Lifshitz formula
for the Casimir--Polder force takes the following form~\cite{6}:\vspace{-6pt}
\begin{equation}
F_{{\rm sub}}(a,T)=-\frac{k_BT}{8a^4}\sum_{l=0}^{\infty}{\vphantom{\sum}}^{\prime}\alpha_l
\int\limits_{\zeta_l}^{\infty}y\,dy\,e^{-y}\left[(2y^2-\zeta_l^2)R_{{\rm TM},l}(y)-
\zeta_l^2R_{{\rm TE},l}(y)\right].
\label{eq1}
\end{equation}\vspace{-6pt}\

\noindent
{In} 
 this equation, the~prime on the summation sign divides the term with $l=0$ by 2 and
\begin{equation}
\alpha_l=\alpha(i\xi_l)=\alpha\left(i\frac{\zeta_lc}{2a}\right)
\label{polarizability}
\end{equation}

\noindent
is the dynamic polarizability of an atom, a~molecule, or~a
nanoparticle taken at a pure imaginary frequency that has the dimension of $\mbox{cm}^3$.

Particular attention should be paid to the quantities $R_{{\rm TM},l}$ and $R_{{\rm TE},l}$ in
(\ref{eq1}); these are the reflection coefficients for the transverse magnetic (TM) and transverse electric (TE),
or equivalently p and s, polarized electromagnetic waves on the planar structure calculated
at the pure imaginary frequencies

\begin{eqnarray}
&&
R_{{\rm TM},l}(y)=R_{\rm TM}(i\xi_l,k_{\bot})=
R_{\rm TM}\left(i\frac{\zeta_lc}{2a},\frac{1}{2a}\sqrt{y^2-\zeta_l^2}\right)
=R_{\rm TM}(i\zeta_l,y),
\nonumber \\[1mm]
&&
R_{{\rm TE},l}(y)=R_{\rm TE}(i\xi_l,k_{\bot})=
R_{\rm TE}\left(i\frac{\zeta_lc}{2a},\frac{1}{2a}\sqrt{y^2-\zeta_l^2}\right)
=R_{\rm TE}(i\zeta_l,y).
\label{eq2}
\end{eqnarray}

In our case, the planar structure is a graphene sheet deposited on a dielectric substrate. The
electromagnetic response of the graphene is described by the polarization tensor in \\ (2 + 1) dimensions,
whereas the material of the substrate reacts to the electromagnetic field through its frequency-dependent
dielectric permittivity. The~form of reflection coefficients in this unusual case was found in~\cite{42}.
Using the dimensionless variables introduced above, these coefficients are provided by~\cite{27}
\begin{eqnarray}
&&
R_{{\rm TM},l}(y)=\frac{\ve_ly(y^2-\zeta_l^2)+\sqrt{y^2+(\ve_l-1)\zeta_l^2}\left[y\tP_{00,l}(y)-
(y^2-\zeta_l^2)\right]} {\ve_ly(y^2-\zeta_l^2)+\sqrt{y^2+(\ve_l-1)\zeta_l^2}\left[y\tP_{00,l}(y)+
(y^2-\zeta_l^2)\right]},
\nonumber \\[1.5mm]
&&
R_{{\rm TE},l}(y)=\frac{(y^2-\zeta_l^2)\left[y-\sqrt{y^2+(\ve_l-1)\zeta_l^2}\right]-\tP_{l}(y)}{(y^2-\zeta_l^2)
\left[y+\sqrt{y^2+(\ve_l-1)\zeta_l^2}\right]+\tP_{l}(y)}.
\label{eq3}
\end{eqnarray}

\noindent
Here, in~analogy with (\ref{polarizability}),
the dielectric permittivity of a substrate is taken at the pure imaginary frequencies
\begin{equation}
\ve_l=\ve(i\xi_l)=\ve\left(i\frac{\zeta_lc}{2a}\right).
\label{permittivity}
\end{equation}

\noindent
The quantity $\tP_{00,l}$ is the 00 component  and $\tP_l$ is
the linear combination of components of
the dimensionless polarization tensor
\begin{equation}
\tP_{kn}=\frac{2a}{\hbar}\Pi_{kn},
\label{tensor}
\end{equation}

\noindent
which are taken at the
pure imaginary frequencies $i\zeta_lc/(2a)$.

The explicit expressions for the quantities $\tP_{00,l}$  and $\tP_l$ for real graphene sheets
characterized by any value of the energy gap $\Delta$ and chemical potential $\mu$ are presented
in~\cite{27}. Here, we use them in the more convenient identical form provided in~\cite{41}.
Thus, the~00 component is
\begin{eqnarray}
&&
\tP_{00,l}(y)=\alpha\frac{y^2-\zeta_l^2}{p_l}\Psi(D_l)+\frac{16\alpha ak_BT}{\Sv\hbar c}
\ln\left[\left(e^{-\frac{\Delta}{2k_BT}}+e^{\frac{\mu}{k_BT}}\right)
\left(e^{-\frac{\Delta}{2k_BT}}+e^{-\frac{\mu}{k_BT}}\right)\right]
\nonumber\\
&&\label{eq4}\\[-2mm]
&&-\frac{4\alpha p_l}{\Sv}\int\limits_{D_l}^{\infty}du \,w_l(u,y)\,{\rm Re}
\frac{p_l-p_lu^2+2i\zeta_l u}{[p_l^2-p_l^2u^2+\Sv(y^2-\zeta_l^2)D_l^2+
2i\zeta_lp_lu]^{1/2}}.
\nonumber
\end{eqnarray}

\noindent
The quantity $\alpha$ in (\ref{eq4}) is the fine structure constant
\begin{equation}
\alpha=\frac{e^2}{\hbar c}\approx\frac{1}{137}.
\label{fine}
\end{equation}

\noindent
It should not be confused with the static polarizability $\alpha_0$ used below.
The function $\Psi(D_l)$ is defined as
\begin{equation}
\Psi(D_l)=2\left[D_l+(1-D_l^2){\rm arctan}\frac{1}{D_l}\right],
\label{eq5}
\end{equation}

\noindent
where
\begin{equation}
D_l\equiv D_l(y)=\frac{2a\Delta}{\hbar cp_l}, \qquad
\qquad
p_l=p_l(y)=\sqrt{\Sv y^2+(1-\Sv)\zeta_l^2}
\label{eq6}
\end{equation}

\noindent
and $\Fv$ is the dimensionless Fermi velocity, $\Fv=v_F/c\approx 1/300$.

As to the function $w_l(u,y)$ entering (\ref{eq4}), it is provided by
\begin{equation}
w_l(u,y)=\frac{1}{e^{B_lu+\frac{\mu}{k_BT}}+1}+
\frac{1}{e^{B_lu-\frac{\mu}{k_BT}}+1},
\label{eq7}
\end{equation}

\noindent
where
\begin{equation}
B_l\equiv B_l(y)=\frac{\hbar cp_l(y)}{4ak_BT}.
\label{eq8}
\end{equation}

Using the same notations, we express the linear combination of the components of the polarization
tensor $\tP_l$ from Equation~(\ref{eq3}) defining the TE reflection coefficient~\cite{27,41}:\vspace{-9pt}

\begin{adjustwidth}{-\extralength}{0cm}
\centering 
\begin{eqnarray}
&&
\tP_{l}(y)=\alpha(y^2-\zeta_l^2)p_l\Psi(D_l)-\frac{16\alpha ak_BT\zeta_l^2}{\Sv\hbar c}
\ln\left[\left(e^{-\frac{\Delta}{2k_BT}}+e^{\frac{\mu}{k_BT}}\right)
\left(e^{-\frac{\Delta}{2k_BT}}+e^{-\frac{\mu}{k_BT}}\right)\right]
\nonumber\\
&&\label{eq9}\\[-2mm]
&&+\frac{4\alpha p_l^2}{\Sv}\int\limits_{D_l}^{\infty}du \,w_l(u,y)\,{\rm Re}
\frac{\zeta_l^2-p_l^2u^2+\Sv(y^2-\zeta_l^2)D_l^2+
2i\zeta_l p_l u}{[p_l^2-p_l^2u^2+\Sv(y^2-\zeta_l^2)D_l^2+
2i\zeta_lp_lu]^{1/2}}.
\nonumber
\end{eqnarray}
\end{adjustwidth}

As is seen from (\ref{eq1}), (\ref{eq3})--(\ref{eq9}),  for~computation of the Casimir--Polder
force acting on some particle on the source side of a graphene-coated substrate, it is desirable to
know the dynamic polarizability of this particle (\ref{polarizability}) and the dielectric
permittivity of a substrate (\ref{permittivity}) within a sufficiently wide frequency region,
along with the energy gap, the chemical potential of the graphene sheet, and the temperature.

It is pertinent to note that the above formalism was developed in the framework of
the Dirac model of graphene~\cite{14,15,16}, which applies when the energy is not too high, i.e.,
$\hbar\omega<3~$eV~\cite{43}. Taking into consideration that the characteristic frequency
of the Casimir--Polder force is provided by $c/(2a)$, which leads to energies of less than 1~eV
at all separations exceeding 100~nm~\cite{6}, it can be concluded that this formalism is
very well adapted for theoretical description of the Casimir--Polder force at large
separations.

\section{Impact of Substrate on the Casimir--Polder Force from a Real Graphene Sheet
at Large~Separations}\label{sec:3}

For a particle interacting with a lonely graphene sheet which is freestanding in vacuum,
it was shown in~\cite{41} that, for separations exceeding some value $a_0$ from 2.3 to
$3.2~\upmu$m, 99\% of the total Casimir--Polder force is provided by the term of (\ref{eq1})
with $l=0$.  This result is valid for various values of the energy gap and chemical
potential of graphene. We start by finding how it is modified by the presence of a~substrate.

For this purpose, we consider the term of (\ref{eq1}) with $l=0$\vspace{-6pt}
\begin{equation}
F_{{\rm sub},0}(a,T)=-\frac{k_BT}{8a^4}\alpha_0
\int\limits_{0}^{\infty}y^3dy\,e^{-y}R_{{\rm TM},0}(y),
\label{eq10}
\end{equation}

\noindent
where $\alpha_0=\alpha(0)$ is the static polarizability of an atom or a nanoparticle.
It is seen that $F_{{\rm sub},0}$ does not depend on the TE reflection~coefficient.

The reflection coefficient $R_{{\rm TM},0}$ is obtained from (\ref{eq3}), again
by setting $l=0$,
\begin{equation}
R_{{\rm TM},0}(y)=\frac{\ve_0y+\tP_{00,0}(y)-y}{\ve_0y+\tP_{00,0}(y)+y}.
\label{eq11}
\end{equation}

When obtaining (\ref{eq11}), it was assumed that
\begin{equation}
\lim_{\zeta\to 0}\left[\zeta^2\ve\left(i\frac{\zeta c}{2a}\right)\right]=0,
\label{eq12}
\end{equation}

\noindent
which is valid for substrates made of both dielectric materials and metals described
by the Drude model. The~value of $\tP_{00,0}(y)$ in (\ref{eq11}) is provided by (\ref{eq4})
with $l=0$ {\cite{27}} 
\begin{eqnarray}
&&
\tP_{00,0}(y)=\frac{\alpha y}{\Fv}\Psi(D_0)+\frac{16\alpha ak_BT}{\Sv\hbar c}
\ln\left[\left(e^{-\frac{\Delta}{2k_BT}}+e^{\frac{\mu}{k_BT}}\right)
\left(e^{-\frac{\Delta}{2k_BT}}+e^{-\frac{\mu}{k_BT}}\right)\right]
\nonumber\\[0mm]
&&\label{eq13}\\[-2mm]
&&-\frac{4\alpha y}{\Fv}\int\limits_{D_0}^{\sqrt{1+D_0^2}}du \,w_0(u,y)\,
\frac{1-u^2}{\sqrt{1-u^2+D_0^2}}.
\nonumber
\end{eqnarray}

\noindent
Here, the~quantity $D_0$ is obtained from (\ref{eq6}) and $B_0$ entering $w_0$ is obtained from (\ref{eq8})
\begin{equation}
D_0=\frac{2a\Delta}{\hbar c\Fv y}\, , \qquad
B_0=\frac{\hbar c\Fv y}{4ak_BT}.
\label{eq14}
\end{equation}

\noindent
{{Note that Equation~(14) of {\cite{41}} 
 for $\tP_{00,0}$ contains a typo: the factor $y^2$ in the first term on the right-hand side of (14) in {\cite{41}} 
 should be replaced with $y$.}}

All numerical computations below are performed for a graphene sheet deposited on
a fused silica glass
(SiO$_2$) substrate. This substrate is typical in graphene technologies
(see, e.g.,~\cite{44,45,46,47}) and
in precision measurements of the Casimir force from graphene~\cite{48,49}. The~optical data for
the complex index of refraction of fused silica glass can be found in~\cite{50} in the wide range
of $\hbar \omega$ extending from 0.0025~eV to 2000~eV. The~obtained dielectric permittivity of
SiO$_2$ along the imaginary frequency axis is well known and has been used in many publications
(see, e.g.,~\cite{6}). It contains the two steps, one of which is due to ionic
polarization and another one is due to electronic polarization. The~static dielectric permittivity
of SiO$_2$ is $\varepsilon_0=3.81$.

Now, we compute the large-separation Casimir--Polder force from a graphene-coated SiO$_2$
substrate using the full Lifshitz formula (\ref{eq1}) and its term with $l=0$ (\ref{eq10})
for various values
of the energy gap and chemical potential. Taking into account that in the presence of
the substrate the region
of large separations starts at separations exceeding a few micrometers, full computations
using (\ref{eq1})
can be safely performed by setting $\alpha_l\approx \alpha_0$ without sacrificing
precision~\cite{6}.
When using the Lifshitz formula (\ref{eq1}), the~computations are performed by
\mbox{(\ref{eq3})--(\ref{eq9}).}
When Equation~(\ref{eq10}) is employed, we use  (\ref{eq11}),  (\ref{eq13}), and~ (\ref{eq14}) in
the computations. In~both cases, we use the value of $\varepsilon_0$ for the SiO$_2$ substrate, and in full
computations we use values of $\varepsilon_l$ with $l\geqslant 1$.

Comparing the obtained results, we find that the full values of the Casimir--Polder force
$F_{\rm{sub}}$
computed by (\ref{eq1}) differ from the force $F_{\rm{sub},0}$ computed by (\ref{eq10}) in less than
1\% at all separations $a\geq a_0=5.6~\upmu$m independently {{of}} the values of $\Delta$ and $\mu$
used in computations (see below for the typical specific values of these parameters). Thus, for~a
graphene sheet deposited on a substrate, the large-separation behavior of the Casimir--Polder force
provided by  $F_{\rm{sub},0}$ is reached at larger separations than for a freestanding graphene sheet
(i.e., from~2.3 to 3.2 $\upmu$m, depending on the values of $\Delta$ and $\mu$ \cite{41}). What
is more, for~a graphene-coated substrate the value of $a_0$ is essentially independent on
$\Delta$ and $\mu$. Note that for an uncoated silica glass plate the large-separation behavior
of the Casimir--Polder force is reached at $a_0=6$ $\upmu$m.

\begin{figure}[H]
\includegraphics[width=3in]{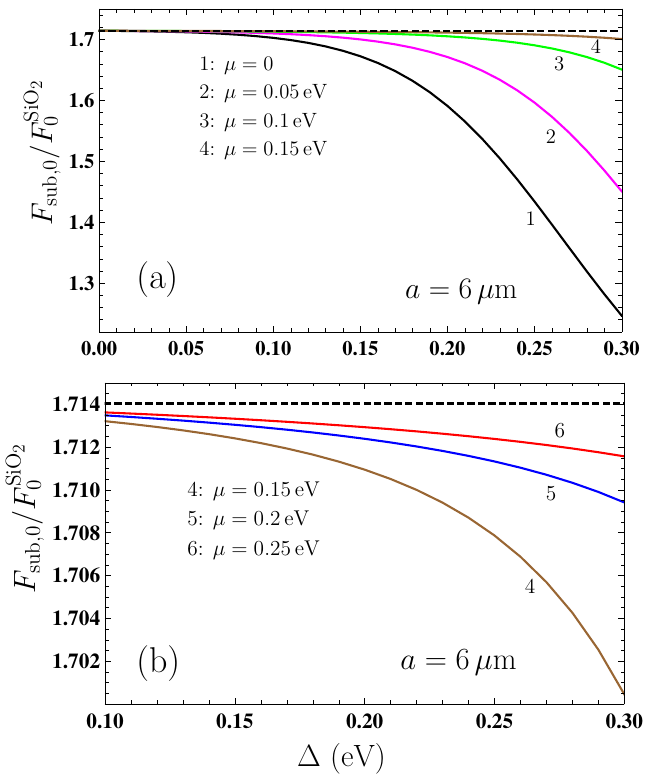}
\caption{\label{fg1}%
{The} 
 ratio of the Casimir--Polder force from the graphene-coated SiO$_2$ substrate
to that from an uncoated plate at the distance of 6 $\upmu$m is shown as the function of the energy gap
(\textbf{a}) by the lines labeled 1, 2, 3, and~4 for the chemical potential of graphene equal to 0, 0.05, 0.1,
and 0.15~eV, respectively, and~(\textbf{b}) by the lines labeled 4, 5, and~6 for the chemical potential equal
to 0.15, 0.2, and~0.25~eV, respectively. The~dashed lines show the ratio of the Casimir--Polder force
from an ideal metal plane to that from an uncoated SiO$_2$ plate.}
\end{figure}

In an effort to determine the impact of graphene coating on the large-separation Casimir--Polder force,
we computed the force $F_{\rm{sub},0}$ from a graphene-coated SiO$_2$ substrate and
$F_0^{\rm{SiO}_2}$ from an uncoated fused silica plate. To~compute the latter quantity, one
should set $\tilde {\Pi}_{00,0}(y)=0$ in (\ref{eq11}). In~Figure~\ref{fg1}a,b the
computational results for the ratio $F_{\rm{sub},0}/F_0^{\rm{SiO}_2}$ at $T$ = 300 K,
$a=6~\upmu$m are shown as a function of the energy gap by (a) the four lines labeled
1, 2, 3, and~4 for $\mu= 0$, 0.05, 0.1, and~0.15~eV, respectively, and (b) the three lines labeled 4, 5, and~6
for $\mu= 0.15$, 0.2, and~0.25~eV, respectively. For~illustrative purposes, the~dashed lines show the value of the
ratio $F_0^{\rm{IM}}/F_0^{\rm{SiO}_2}$ when the graphene-coated substrate is replaced with an
ideal metal~plane.

As seen in Figure~\ref{fg1}a,b, the~impact of graphene coating on the large-separation
Casimir--Polder force increases with increasing chemical potential and decreasing energy gap of
the graphene. As~a result, the~fused silica substrate coated by a graphene sheet with $\Delta = 0.1$~eV
and $\mu = 0.25$~eV produces almost the same Casimir--Polder force at $a = 6~\upmu$m separation
as an ideal metal~plane.

It is well known that at separations above 6~$\upmu$m the Casimir--Polder force from an ideal metal
plane takes the so-called classical form, which does not depend on either $\hbar$ or $c$ \cite{6}:
\begin{equation}
F_0^{\rm IM}(a,T)=-\frac{3k_BT}{4a^4}\alpha_0.
\label{eq15}
\end{equation}

In order to find how the large-separation Casimir--Polder force from a graphene-coated SiO$_2$
substrate approaches this classical limit depending on the values of $\Delta$ and $\mu$, we can calculate
the relative difference
\begin{equation}
\delta F_{{\rm sub},0}(a,T)=\frac{F_{{\rm sub},0}(a,T)- F_0^{\rm IM}(a,T)}{F_0^{\rm IM}(a,T)}.
\label{eq16}
\end{equation}

\begin{figure}[H]
\includegraphics[width=3in]{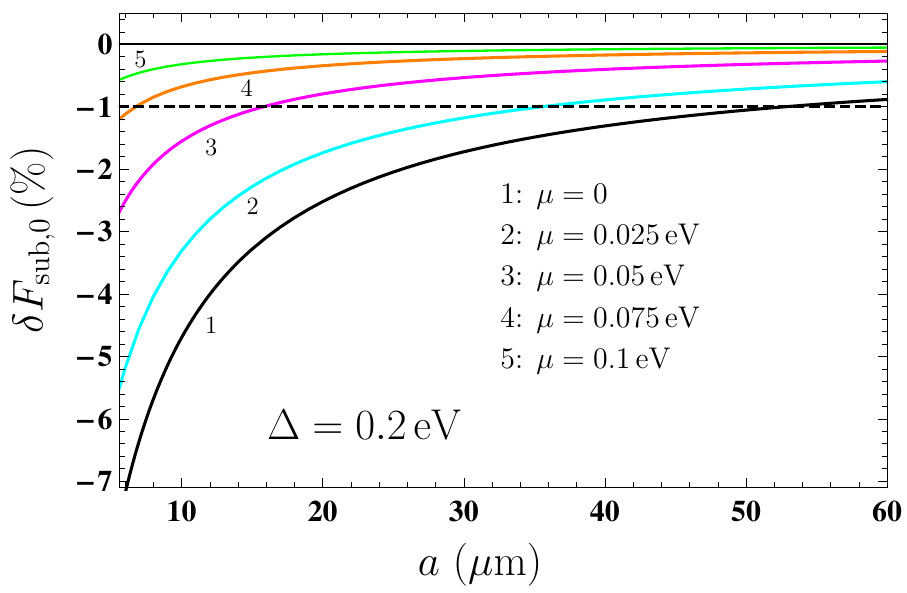}
\caption{\label{fg2}%
{The} 
 relative difference between the large-separation Casimir--Polder forces from the
graphene-coated SiO$_2$ substrate and an ideal metal plane for the energy gap of graphene
equal to 0.2~eV at $T$ = 300~K is shown as the function of separation by the lines labeled
 1, 2, 3, 4, and~5 for the chemical potential of graphene equal to 0, 0.025, 0.05, 0.075,
and 0.1~eV, respectively. The~dashed line restricts the area of the figure plane where
$\delta F_{{\rm sub},0}\leqslant 1\%$. }
\end{figure}

In Figure~\ref{fg2}, the~computational results for $\delta F_{\rm{sub},0}$ at $T$ = 300 K are shown for a
graphene coating with $\Delta = 0.2$~eV as a function of separation by the five lines labeled
1, 2, 3, 4, and~5 for $\mu = 0$, 0.025, 0.05, 0.075, and~0.1~eV, respectively. The~dashed line shows the
lower boundary of the figure plane domain where the relative deviation between  $F_{\rm{sub},0}$ and
$F_0^{\rm{IM}}$ is less than 1\%.

As is seen in Figure~\ref{fg2}, for~a graphene coating with $\mu =0.1$~eV (line 5), the
relative deviation
(\ref{eq16}) remains within 1\% over the entire separation region $a > 5.6 ~\upmu$m.
As for the
graphene coating with smaller $\mu = 0.075$, 0.05, 0.025, and~0~eV, the~Casimir--Polder
force takes the
classical form at larger separations equal to 6.5, 15.5, 35.5, and~53~$\upmu$m, respectively.
These values
are only slightly larger than for a freestanding graphene sheet~\cite{41}.
From Figure~\ref{fg2}, it can be
concluded that the Casimir--Polder
force from a graphene-coated
substrate takes the classical form at larger separations for graphene with lower chemical potential.

\begin{figure}[H]
\includegraphics[width=3.5in]{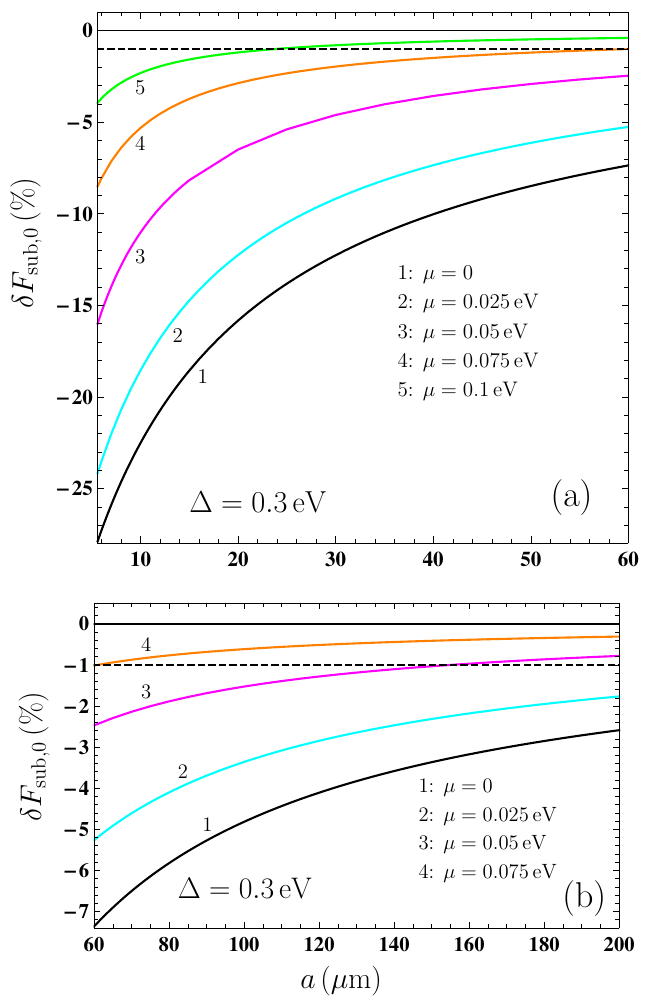}
\caption{\label{fg3}%
{The} 
 relative difference between the large-separation Casimir--Polder forces from the
graphene-coated SiO$_2$ substrate and an ideal metal plane for the energy gap of graphene
equal to 0.3~eV at $T$ = 300~K is shown as the function of separation (\textbf{a}) by the lines labeled
1, 2, 3, 4, and~5 for the chemical potential of graphene equal to 0, 0.025, 0.05, 0.075,
and 0.1~eV, respectively, in the separation region from 5.6 $\upmu$m to 60 $\upmu$m and (\textbf{b}) by the lines labeled
1, 2, 3, and~4 for the chemical potential of graphene equal to 0, 0.025, 0.05, and~0.075~eV, respectively, in the
separation region from 60 $\upmu$m to 200 $\upmu$m. The~dashed line restricts the areas of the figure plane
where $\delta F_{{\rm sub},0}\leqslant 1\%$. }
\end{figure}

In fact, for a graphene sheet deposited on a substrate, the initiation of an energy gap is almost
unavoidable~\cite{14}. Thus, an~energy gap equal to 0.3~eV is rather typical~\mbox{\cite{48,49}.}
To cover this case, in~Figure~\ref{fg3}a,b we present the computational results for
$\delta F_{\rm{sub},0}$ at \mbox{$T$ = 300 K} for a graphene coating with $\Delta = 0.3$~eV. These
results are again shown as the function of separation (a) by the five lines labeled 1, 2, 3, 4, and~5
for $\mu = 0$, 0.025, 0.05, 0.075, and~0.1~eV, respectively, over the separation interval from 5.6 to 60~$\upmu$m
and (b) by the four lines labeled 1, 2, 3, and~4 for $\mu = 0$, 0.025, 0.05,
and 0.075~eV, respectively, over the separation interval from 60 to 200~$\upmu$m. The~dashed lines in
Figure~\ref{fg3}a,b again enclose the figure domain where $\delta F_{\rm{sub},0}$ is less than 1\%.

As is seen in  Figure~\ref{fg3}a, only for a graphene coating with $\mu = 0.1$~eV (line 5) does the
Casimir--Polder force take the classical form in the field of this figure (for $a > 24.5~\upmu$m). Thus,
with increasing energy gap the classical form of the force is reached at larger separations. From
Figure~\ref{fg3}b, it can be seen that for graphene sheets with $\mu = 0.075$ and 0.05~eV the
Casimir--Polder force becomes classical starting from 63 and 157~$\upmu$m, respectively (lines 4 and 3).
As for the graphene coating with $\mu = 0.025$ and 0~eV, the~corresponding Casimir--Polder force does
not take the classical form in the field of this figure up to $a = 200~\upmu$m. Calculations show that
the Casimir--Polder force from the graphene-coated substrate with $\Delta = 0.3$~eV, $\mu = 0.025$
and 0~eV becomes classical only at separations of 363 and 550~$\upmu$m, respectively. By~and large,
Figures~\ref{fg2} and \ref{fg3} demonstrate that with increasing $\mu$ the classical form
of the Casimir--Polder force from the graphene-coated substrate is reached at shorter and
with increasing $\Delta$ at larger~separations.

\section{Analytic Asymptotic Results Confronted with Numerical Computations in the
Presence of~Substrate}\label{sec:4}

In this section, we obtain the analytic asymptotic expressions $F_0^{\rm as}$ for the
large-separation Casimir--Polder force from a graphene-coated substrate $F_{{\rm sub},0}$
and find the measure of agreement between $F_0^{\rm as}$ and the numerically computed
$F_{{\rm sub},0}$ for different values of the energy gap and chemical potential of a
graphene~sheet.

The asymptotic expression sought here is valid under the following condition:
\begin{equation}
\frac{2ak_BT}{\Fv\hbar c}\gg 1,
\label{eq17}
\end{equation}

\noindent
which is satisfied with a large safety margin at $T=300~$K, $a>0.2~\upmu$m (in fact, as
was shown in Section~3, we consider separation distances exceeding $5.6~\upmu$m).

We consider the large-separation Casimir--Polder force (\ref{eq10}) with the reflection
coefficient (\ref{eq11}) under the condition  (\ref{eq17}). This reflection coefficient
can be rearranged to
\begin{equation}
R_{\rm TM}(0,y)=1-\frac{2y}{\tP_{00,0}(y)+(\ve_0+1)y}.
\label{eq18}
\end{equation}

According to (\ref{eq13}), the~polarization tensor is of the order of parameter (\ref{eq17});
thus, $\tP_{00,0}(y)\gg 1$. What is more, the~main contribution to the integral
(\ref{eq10}) is provided by $y\sim 1$. Because~of these conditions, we can replace $y$ with
unity in the denominator of (\ref{eq18}) and neglect $(\ve_0+1)$ in comparison
with $\tP_{00,0}(1)$. Therefore, (\ref{eq18}) becomes\vspace{-6pt}
\begin{equation}
R_{\rm TM}(0,y)\approx 1-\frac{2y}{\tP_{00,0}(1)},
\label{eq19}
\end{equation}

\noindent
i.e., the~TM reflection coefficient takes exactly the same approximate form as was found
earlier for the freestanding graphene sheet~\cite{41}.

After substitution of (\ref{eq19}) in (\ref{eq10}) and integration, the~desired asymptotic
expression takes the same form as in~\cite{41}\vspace{-6pt}
\begin{equation}
F_{{\rm sub},0}^{\rm as}(a,T)=F_0^{\rm as}(a,T)=F_0^{\rm IM}(a,T)\left[1-
\frac{8}{\tP_{00,0}(1)}\right],
\label{eq20}
\end{equation}

\noindent
where $F_0^{\rm IM}(a,T)$ is defined in (\ref{eq15}).

The expression (\ref{eq20}) should be supplemented by the approximate expression for
$\tP_{00,0}(1)$ found in~\cite{41} under the condition (\ref{eq17})
\begin{eqnarray}
&&
\tP_{00,0}(1)\approx \frac{16\alpha a k_BT}{\Sv\hbar c}\left[\ln\left(
4\cosh\frac{\Delta+2\mu}{4k_BT}\cosh\frac{\Delta-2\mu}{4k_BT}\right)\right.
\nonumber \\[1mm]
&&~~~~~~~~\left.
-\frac{\Delta}{4k_BT}\left(\tanh\frac{\Delta+2\mu}{4k_BT}+
\tanh\frac{\Delta-2\mu}{4k_BT}\right)\right].
\label{eq21}
\end{eqnarray}

We are coming now to a detailed comparison between the asymptotic expression for
the Casimir--Polder force $F_0^{\rm as}$ provided by (\ref{eq20}) and (\ref{eq21})
and the numerically computed large-separation force $F_{{\rm sub},0}(a,T)$
from a graphene-coated substrate using Equations~(\ref{eq10}), (\ref{eq11}),
and (\ref{eq13}). This comparison can be made by considering the ratio
$F_{{\rm sub},0}/F_0^{\rm as}$ for different values of the graphene parameters.
All computations are again performed at room temperature $T=300~$K.

\begin{figure}[H]
\includegraphics[width=4in]{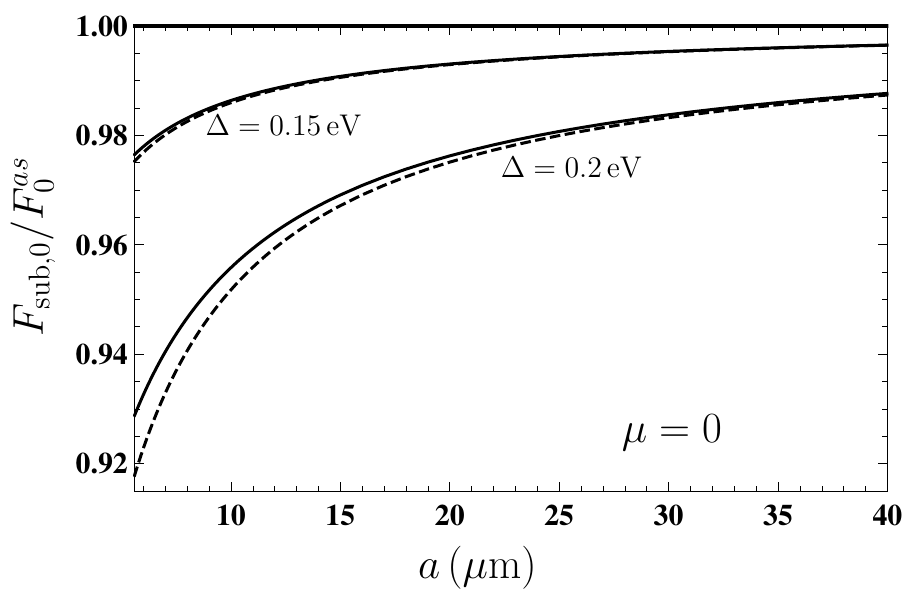}
\caption{\label{fg4}%
{The} 
 ratio of the numerically computed large-separation Casimir--Polder force from the
graphene-coated SiO$_2$ substrate to its asymptotic value for the zero chemical potential of graphene
at $T$ = 300~K is shown as the function of separation by the top and bottom solid lines for the energy
gap equal to 0.15 and 0.2~eV, respectively. The similar ratio for the freestanding graphene sheets with
the same parameters is shown by the dashed lines. }
\end{figure}

First, we consider an undoped graphene coating with $\mu=0$ and compute the ratio
$F_{{\rm sub},0}/F_0^{\rm as}$ for two different values of the energy gap
$\Delta=0.15~$eV and 0.2~eV. The~computational results are shown in Figure~\ref{fg4}
as the function of separation by the top and bottom solid lines, respectively.
For comparison purposes, the~ratio $F_0/F_0^{\rm as}$ for a freestanding graphene
sheet (i.e., with~no substrate) with the same values of $\Delta$ is plotted by the
dashed lines.\vspace{-6pt}

As seen in Figure~\ref{fg4}, for~a graphene coating with $\mu=0$, $\Delta=0.15~$eV,
the asymptotic results reproduce the results of the numerical computations with better than
1\% accuracy at separations exceeding $14.5~\upmu$m. With~increasing energy gap, the
accuracy of the asymptotic expressions becomes lower. Thus, for~a graphene coating
with $\mu=0$, $\Delta=0.2~$eV, the asymptotic expression is accurate within 1\%
starting from $25~\upmu$m separation. From~Figure~\ref{fg4}, it can be seen that
in spite of the fact that the asymptotic expression used does not depend on the
dielectric permittivity of a substrate it reproduces the results of numerical
computations somewhat better than for a freestanding graphene sheet. This is
illustrated by the solid lines, which lie slightly above the dashed~ones.

\begin{figure}[H]
\includegraphics[width=4in]{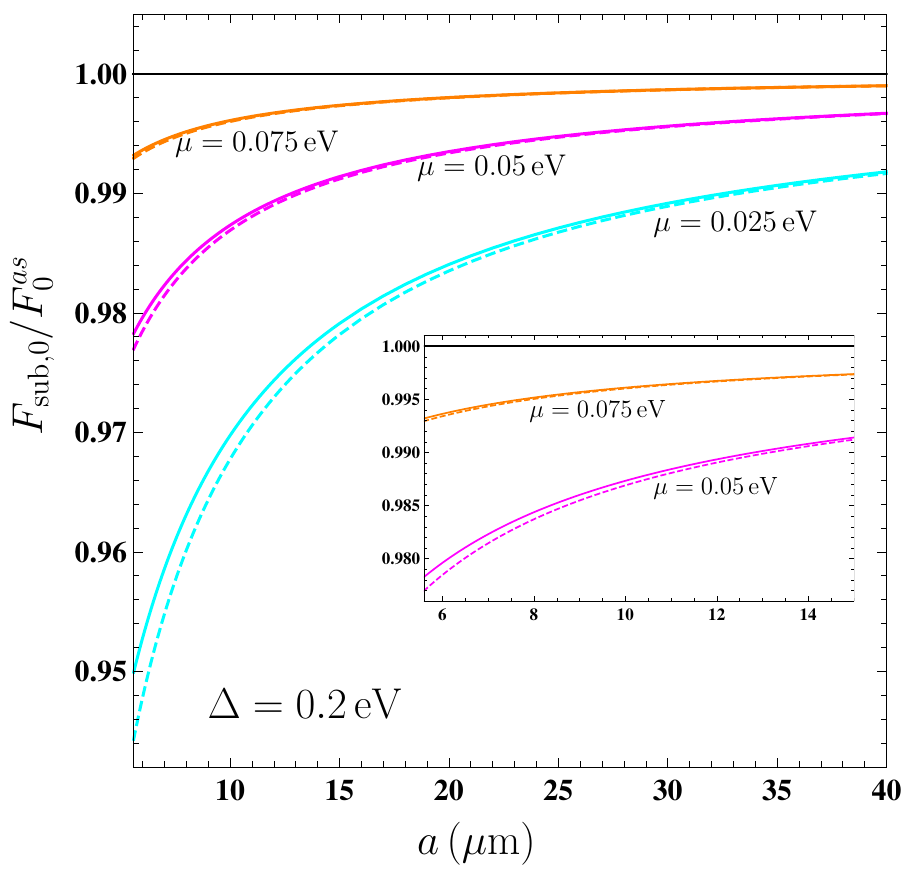}
\caption{\label{fg5}%
{The} 
 ratio of the numerically computed large-separation Casimir--Polder force from the
graphene-coated SiO$_2$ substrate to its asymptotic value for the energy gap of graphene
equal to 0.2~eV at $T$ = 300~K is shown as the function of separation by the three solid lines
counted from bottom to top for the chemical potential equal to 0.025, 0.05, and~0.075~eV,
respectively. The~dashed lines show similar ratio for a freestanding graphene sheet with the
same parameters. The~region of short separations is reproduced in the inset on an enlarged scale
for graphene sheets with a chemical potential of 0.05~eV (bottom) and 0.075~eV (top). }
\end{figure}

Next, we consider a graphene coating with a reasonably large energy
gap $\Delta = 0.2$~eV and
consider the relationship between the asymptotic and numerical results for a large-separation
Casimir--Polder force from a graphene-coated substrate for different values of the chemical
potential. The~computational results for the ratio $F_{\rm{sub},0}/F_0^{\rm{as}}$ in this case
at \mbox{$T$ = 300 K} are presented in Figure~\ref{fg5} by the three solid lines from bottom to top for the
chemical potential equal to $\mu = 0.025$, 0.05, and~0.075~eV, respectively.
The dashed lines show
the ratio $F_{0}/F_0^{\rm{as}}$ for the freestanding graphene sheets with the same values of $\mu$.
In the inset, the~two pairs of solid and dashed lines for $\mu = 0.05$ and 0.075~eV are reproduced
on an enlarged scale within a narrower separation region for better~visualization.

From Figure~\ref{fg5}, it can be seen that with increasing $\mu$ the agreement between the asymptotic
and numerically computed Casimir--Polder forces from the graphene-coated substrate becomes better.
Thus, if~for a graphene coating with $\mu = 0.025$~eV the 1\% agreement is reached only at
separations exceeding 41~$\upmu$m, for~$\mu = 0.05$~eV this measure of agreement is observed
at $a > 13.5~\upmu$m. As~to the graphene coating with $\mu = 0.075$~eV, the~better than 1\%
agreement between the asymptotic and numerical results holds over the entire range of large
separations $a \geq 5.6~\upmu$m. According to  Figure~\ref{fg5} (see the inset as well), the solid
lines lie above the dashed ones for all $\mu$, i.e.,~the asymptotic expression (\ref{eq20}) and
(\ref{eq21}) is slightly more exact in the case of graphene-coated substrates than for
freestanding graphene~sheets.

\begin{figure}[H]
\includegraphics[width=3.5in]{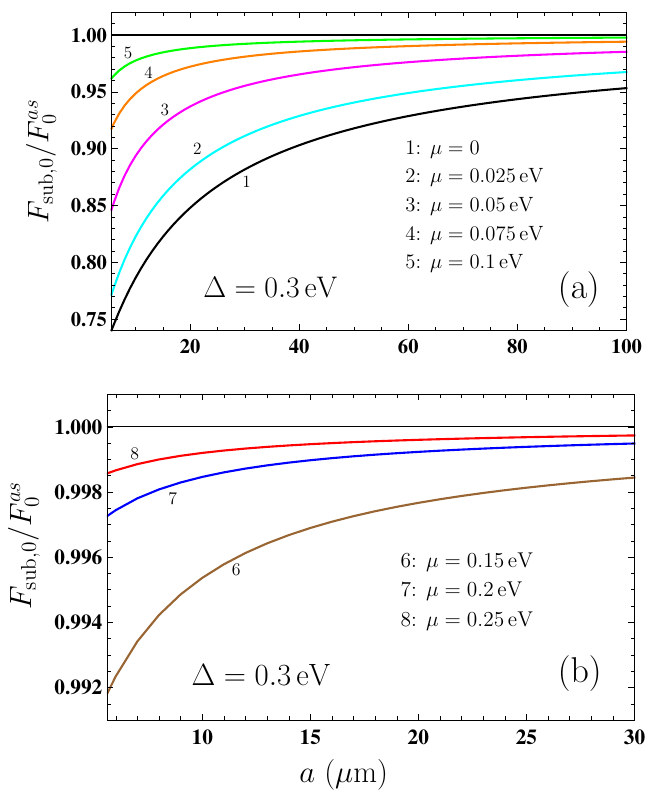}
\caption{\label{fg6}%
{The} 
 ratio of the numerically computed large-separation Casimir--Polder force from the
graphene-coated SiO$_2$ substrate to its asymptotic value for the energy gap of graphene
equal to 0.3~eV at $T$ = 300~K is shown as the function of separation by the solid lines
counted from bottom to top (\textbf{a})~labeled 1, 2, 3, 4, and~5 for the chemical potential of
graphene equal to 0, 0.025, 0.05, 0.075, and~0.1~eV, respectively and (\textbf{b}) labeled 6, 7, and~8
for the chemical potential equal to 0.15, 0.2, and~0.25~eV, respectively. }
\end{figure}

At the end of this section, we consider the case of a graphene coating with a larger energy
gap $\Delta = 0.3$~eV, such as that used in the experiment measuring the Casimir force
from a graphene-coated substrate in~\cite{48,49}. Here, we calculate the ratio
$F_{\rm{sub},0}/F_0^{\rm{as}}$ over the wider range of $\mu$ up to $\mu = 0.25$~eV
(the latter value was measured for the sample used in the experiment~\cite{48,49}). In
Figure~\ref{fg6}a,b, we present the computational results for this ratio at $T = 300$~K
as a function of separation. The~obtained results are shown by (a) the five lines labeled
1, 2, 3, 4, and~5 for $\mu = 0$, 0.025, 0.05, 0.075, and~0.1~eV, respectively, over the separation
interval from 5.6~$\upmu$m to 100~$\upmu$m and (b) by the three lines labeled 6, 7, and~8 for
$\mu = 0.15$, 0.2, and~0.25~eV, respectively, over the interval from 5.6 to 30~$\upmu$m.

As seen in Figure~\ref{fg6}a, for lines 1, 2, and~3 (i.e., for~the graphene coatings
with  $\mu = 0$, 0.025, and~0.05~eV, respectively) the 1\% agreement between the asymptotic and
numerical results is not reached up to the separation of 100 $\upmu$m. As~for the lines 4 and 5
($\mu = 0.075$, and~0.1~eV, respectively) the 1\% agreement is reached at separation
distances exceeding approximately 60 and 24~$\upmu$m. Thus, the~agreement between
the asymptotic and numerical results again becomes better with increasing value of the
chemical~potential.

From Figure~\ref{fg6}b, an inference can be drawn that sufficiently large values of $\mu$ can
fully compensate the negative role played by large $\Delta$ in an agreement between the
asymptotic and numerical values of the large-separation Casimir--Polder force from
graphene-coated substrates. According to Figure~\ref{fg6}b, for~all the three graphene
coatings with $\mu = 0.15$, 0.2, and~0.25~eV the asymptotic results are within 1\%
agreement with the results of numerical computations at all separations exceeding the border
of the large-separation region equal to 5.6~$\upmu$m. These results make possible the reliable
use of the analytic asymptotic expression for the Casimir--Polder force from graphene-coated
substrates with the proper combination of the values of $\Delta$ and $\mu$.

\section{Discussion}\label{sec:5}

The Casimir--Polder force considered in this paper is both a quantum and relativistic
phenomenon which has no explanation on the basis of classical physics even in the case
of atoms and nanoparticles interacting with conventional metallic and dielectric materials.
It has been commonly believed, however, that at separations of a few micrometers the
Casimir--Polder force takes its limiting form of large separations where it becomes
classical and no longer depends on either the Planck constant or the speed of light.

The new two-dimensional material called graphene possesses many unusual properties.
One of these properties is a giant thermal effect in the Casimir force at short
separations, predicted in~\cite{51} and experimentally discovered in~\cite{48,49}.
Another unusual property, first discussed in~\cite{41} using the simplified example
of a freestanding graphene sheet in a vacuum, is that while the Casimir--Polder force between
atoms, nanoparticles, and graphene reaches its limiting form of large separations at
distances of 2--3 $\upmu$m,~this form may significantly deviate from being classical
at distances up to tens of micrometers depending on the values of the energy gap and the chemical
potential of~graphene.

Here, we demonstrate that this unusual property is fully preserved in the case of
graphene-coated dielectric substrates, which is of great physical significance. It is
shown that although the limit of large separations for the Casimir--Polder force is
reached at larger separations approximately equal to 5.6 $\upmu$m in this case, the
classical limit may be reached at tens and even at hundreds of micrometers if the
energy gap of the graphene coating is rather high and its chemical potential is rather
low. As~long as the classical limit is not reached, the~Casimir--Polder force from a
graphene-coated substrate depends on both the Planck constant and the speed
of light, i.e., it~remains a quantum and relativistic phenomenon. This effect is
most pronounced in the application region of the analytic asymptotic expression in
(\ref{eq20}) and (\ref{eq21}), which depends on both $\hbar$ and $c$. By~manufacturing
graphene sheets with a small energy gap and sufficiently large chemical potential,
it is possible to obtain the large-separation Casimir--Polder force from the graphene-coated
substrate that is close to that from an ideal metal~plane.

\section{Conclusions}\label{sec:6}

In the foregoing, we have analyzed the behavior of the Casimir--Polder force acting
on atoms or nanoparticles from a graphene-coated substrate spaced at a separation of
a few micrometers. We have shown that the limit of large separations is reached
in this case at a distance of approximately 5.6 $\upmu$m, which is almost
independent on the energy gap and chemical potential of a graphene sheet.
Although the limit of large separations is reached at a distance similar to that for
conventional dielectric and metallic materials, we demonstrate that for
graphene-coated substrates the Casimir--Polder force may attain the classical regime
at much larger distances depending on the values of the energy gap and chemical
potential of the graphene~coating.

In addition, we find the analytic asymptotic expression for the Casimir--Polder force
between atoms (nanoparticles) and graphene-coated substrates at large separations
and determine the region of its applicability. The~obtained expressions allow simple
calculation of the Casimir--Polder force for substrates made of various materials and
graphene coatings with any energy gap and chemical~potential values.

The above results are derived in the framework of rigorous fundamental theory
using the polarization tensor of graphene in the application region of the Dirac model.
The results can be used in numerous applications of graphene in nanotechnology and
bioelectronics, including such areas as field-effect transistors, interaction with lipid
membranes, and graphene--semiconductor~nanocomposites.

\vspace{6pt}

\funding{{G.L.K.} 
 was partially funded by the
Ministry of Science and Higher Education of the Russian Federation
(``The World-Class Research Center: Advanced Digital Technologies'',
contract No. 075-15-2022-311 dated 20 April 2022). The~research
of V.M.M. was partially carried out in accordance with the Strategic
Academic Leadership Program ``Priority 2030'' of Kazan Federal
University. }

\begin{adjustwidth}{-\extralength}{0cm}

\reftitle{References}

\PublishersNote{}
\end{adjustwidth}
\end{document}